\documentclass[aps,twocolumn,english,showpacs,floatfix]{revtex4}
\usepackage[T1]{fontenc}
\usepackage[latin1]{inputenc}
\usepackage{float}
\usepackage{graphicx}
\usepackage{amssymb}

\begin{document}

\title{Superscars in billiards -- A model for doorway states in quantum spectra}

\author{S.~\AA berg$^1$, T.~Guhr$^{1,2}$, M.~Miski--Oglu$^3$ and A.~Richter$^3$}
\affiliation{$^1$
         Matematisk Fysik, LTH, Lunds Universitet,
         Box 118, 22100 Lund, Sweden\\
         $^2$
         Fachbereich Physik, Universit\"at Duisburg--Essen,
         Lotharstrasse 1,  47057 Duisburg, Germany\\
         $^3$
         Institut f\"ur Kernphysik, Technische Universit\"at Darmstadt,
         Schlossgartenstrasse 9, Darmstadt, Germany}

\date{\today{}}

\begin{abstract}
  In a unifying way, the doorway mechanism explains spectral
  properties in a rich variety of open mesoscopic quantum systems,
  ranging from atoms to nuclei.  A distinct state and a background of
  other states couple to each other which sensitively affects the
  strength function. The recently measured superscars in the barrier
  billiard provide an ideal model for an in--depth investigation of
  this mechanism.  We introduce two new statistical observables, the
  full distribution of the maximum coupling coefficient to the doorway
  and directed spatial correlators. Using Random Matrix Theory and
  random plane waves, we obtain a consistent understanding of the
  experimental data.
\end{abstract}



\maketitle

Strength function phenomena \cite{BM1} in open mesoscopic quantum
systems are a central object of study in atomic and molecular physics
as well as in atomic clusters, quantum dots, and in nuclear physics .
Often there is a somehow ``distinct'' and ``simple'' excitation whose
amplitude is spread over many ``complicated'' states.  The distinct
state thus acts as a ``doorway'' to the (usually chaotic) background
of the complicated states~\cite{BM1,Zelev1997}. Prime examples are
Isobaric Analog States and multipole Giant Resonances (GR) in nuclear
physics. The strength function is typically of Breit--Wigner (BW)
shape with a characteristic spreading
width~\cite{Harney1986,Zelev1996,Aberg2004,Shev2004}. For examples
from molecules and metal clusters, see
\cite{Guhr1990,Kawata2000,Husein2000}.

\begin{figure}
 \includegraphics[width=\linewidth]{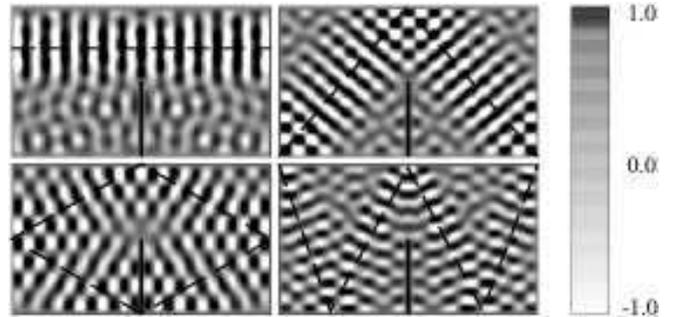}
 \caption{ Examples for measured superscars induced by the barrier and
   concentrated along the indicated classical periodic orbits (dashed
   lines). They are members of four different families. Top row:
   horizontal Bouncing Ball {\sf BB} and Inverted {\sf V} superscars;
   bottom row: Diamond {\sf D} and {\sf W} superscars. The gray level
   indicates the value of the wave function (black: highest positive
   and white: most negative value; see \cite{Bogomol2006}).}
\label{fig1}
\end{figure}

Quantum billiards can be realized experimentally by flat microwave
resonators \cite{StoeckmannBuch2000}.  To study the doorway mechanism
in detail, we use a microwave billiard of rectangular shape with a
thin barrier inside, see Fig.~\ref{fig1}.  The electric field strength
distribution corresponding to the quantum wave function is
reconstructed from the measured intensities. Certain wave functions of
this pseudointegrable billiard possess unique structures called
``superscars''~\cite{Bogomol2004,Bogomol2006}.  These are scarring
wave functions related to families of neutrally stable classical
periodic orbits. Four examples of measured superscars are shown in
Fig.~\ref{fig1}. Unlike ordinary scars \cite{Heller1984}, which are
localized around a single unstable periodic orbit, they do not
disappear at large quantum numbers. They are embedded into, but
clearly distinct from, a large number of nonscarred wave functions. We
will demonstrate that the superscars act as doorways to the background
of the nonscarred wave functions. Our perfect control over the
experimental observables allows us an in--depth study of the doorway
mechanism which can presently not be accomplished in traditional
quantum systems.

First, we briefly compile the necessary information on measured
superscars in the barrier billiard. Second, we introduce the maximal
coupling coefficient as a new observable and use Random Matrix Theory
(RMT)~\cite{Guhr1998} to model its distribution.  Third, we introduce
directed spatial correlators as another new observable and model them
by extending Berry's random wave ansatz~\cite{Berry1977}.

\begin{figure}
 \includegraphics[width=\linewidth]{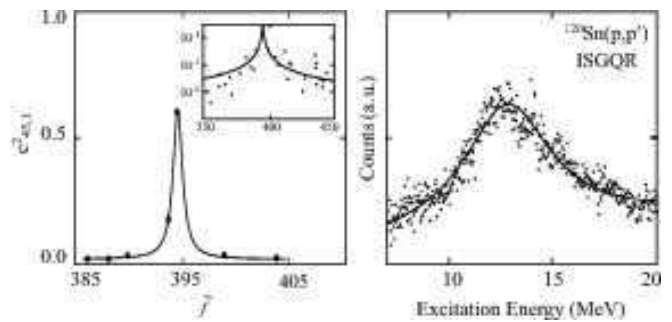}
 \caption{ Doorway strength functions. Left: overlap between constructed superscar state of the
   {\sf V} family with $m=45$, $n=1$ and the measured wave functions
   versus rescaled frequency. The solid curve is a BW
   function. The inset shows the overlap on a logarithmic scale over a
   large frequency interval. Right: spectrum of the $(p,p')$ reaction
   at 200 MeV on $^{120}$Sn in the region of the Isoscalar Giant
   Quadrupole Resonance (ISGQR) as an example for a nuclear doorway
   state \cite{Shev2004}. The solid curve is a BW function fitted to the data.}
\label{fig2}
\end{figure}

As Fig.~\ref{fig1} shows, the superscars with a clear wave function
structure relate to particular classical periodic orbit.  The
superscars form a family which lives within an infinitely long
Periodic Orbit Channel (POC). Due to diffraction on the tip of the
barrier, the amplitude of the scarred wave function tends to zero
along the POC boundary. Thus, the superscarred wave function can be
approximated by a constructed superscar state, defined as an
eigenfunction $\Psi_{m,n}^{\sf (F)}(\vec{r})$ in the infinitely long
POC~\cite{Bogomol2004,Bogomol2006}.  Here ${\sf F}$$\in$$\{ {\sf
  BB,V,D,W} \}$ stands for the superscar families, as defined in
Fig.~\ref{fig1}, and $(m,n)$ are the numbers of wave maxima along and
perpendicular to the POC. A measured state $\Psi_{\tilde{f}}(\vec{r})$
at (rescaled) frequency $\tilde{f}$ in the barrier billiard has an
overlap,
\begin{equation}
c_{m,n}=\langle \Psi_{m,n}^{\sf (F)}|\Psi_{\tilde{f}}\rangle \ , 
\label{overlap}
\end{equation}
with the constructed superscars~\cite{Bogomol2004,Bogomol2006}.  As an
example, the distribution of the overlaps with a constructed {\sf V}
superscar state with quantum numbers $(m,n)=(45,1)$ is depicted in
Fig.~\ref{fig2}. The superscar strength spreads into a neighboring
nonscarring background states following a BW shape with the main
strength concentrated in a few states. This nicely confirms our
doorway interpretation. For comparison, a nuclear GR doorway is also
plotted in Fig.~\ref{fig2}. Here, the number of background states ---
reflected in the fluctuations around its BW shape --- is much larger
than in the barrier billiard.  According to the Brink-Axel hypothesis
\cite{Brink1955,Axel1962} a GR excitation builds upon every nuclear
state. Similarly, a superscar doorway state exists for each value of
${\sf F}$, ${m}$ and $n$.

We now set up a random matrix model in the spirit of models in nuclear
physics~\cite{BM1,Guhr1998}. The total Hamiltonian reads
$\hat{H}=\hat{H}_s + \hat{H}_b + \hat{V}$. Here, $\hat{H}_s$ and
$\hat{H}_b$ describe doorway states and background states,
respectively, and $\hat{V}$ couples the two classes of states. The
eigenequations for the uncoupled Hamiltonians are
$\hat{H}_s|s\rangle=e_s|s\rangle$ and
$\hat{H}_b|b\rangle=e_b|b\rangle$.  For the matrix elements of the
interaction, we make the assumptions $\langle
s|\hat{V}|s'\rangle=\langle b|\hat{V}|b'\rangle=0$ and $\langle
b|\hat{V}|s\rangle=v_{bs}$ for any $s$, $s'$, $b$, $b'$.  We interpret
the constructed superscars, $\Psi^{{\sf (F)}}_{m,n}(\vec{r})$ for a
given family {\sf F} but with different $(m,n)$ as the doorway states
$s$.  Due to the interaction $\hat{V}$ the doorway state is not an
eigenstate of the Hamiltonian $\hat{H}$. We assume that the
interaction matrix elements, $v_{bs}=v_{sb}$, are Gaussian distributed
random variables with zero mean and variance $v^2$. Importantly, the
parameter governing the physics is $v/d$, where $d$ is the mean level
spacing of the background states~\cite{BM1,Guhr1998}.  Since only a
few states carry superscar strength with given values of $(m,n)$, and
superscar states with different $(m,n)$ are assumed to not mix, it is
sufficient to consider only {\em one} superscar state, $s$, coupled to
$N$ background states, $b$, where $N$ is large. To resemble the
experiment, we include $N$=294 background states. As the barrier
billiard is pseudointegrable, the spacings between the eigenstates are
semi--Poisson distributed~\cite{Bogomol1999}.  We thus generate such
an ensemble of $N+1$ states.  The doorway state is chosen as the
middle state and interacts with the surrounding $N$ states.  For each
realization, energies and wave functions are numerically obtained, and
the mixture of the superscar with the surrounding states is
calculated. We then extract $v/d$ for each superscar family.
The full problem $\hat{H}|n\rangle=E_n|n\rangle$ is solved by the
exact implicit equation
\begin{equation}
E_n=e_s-\sum_{b=1}^N\frac{v^2_{bs}}{e_b-E_n} \ ,
\end{equation}
and the wave functions are given by
\begin{equation}
|n\rangle=c_s\left(|s\rangle-\sum_{b=1}^N\frac{v_{bs}}{e_b-E_n}|b\rangle\right) \ .
\end{equation}
The superscar coupling of each eigenstate is therefore
\begin{equation}
c_s(n)=\left(1+\sum_{b=1}^N\frac{v_{bs}^2}{(e_b-E_n)^2}\right)^{-1/2} \ .
\label{cs}
\end{equation}
The superscar strength $c_s^2$ over the different eigenstates
$|n\rangle$ is BW distributed~\cite{BM1} with spreading
width $\Gamma^{\downarrow}=2\pi v^2/d$,
i.e.~$\Gamma^{\downarrow}/d=2\pi (v/d)^2$.

\begin{figure}[!h]
\includegraphics[width=\linewidth]{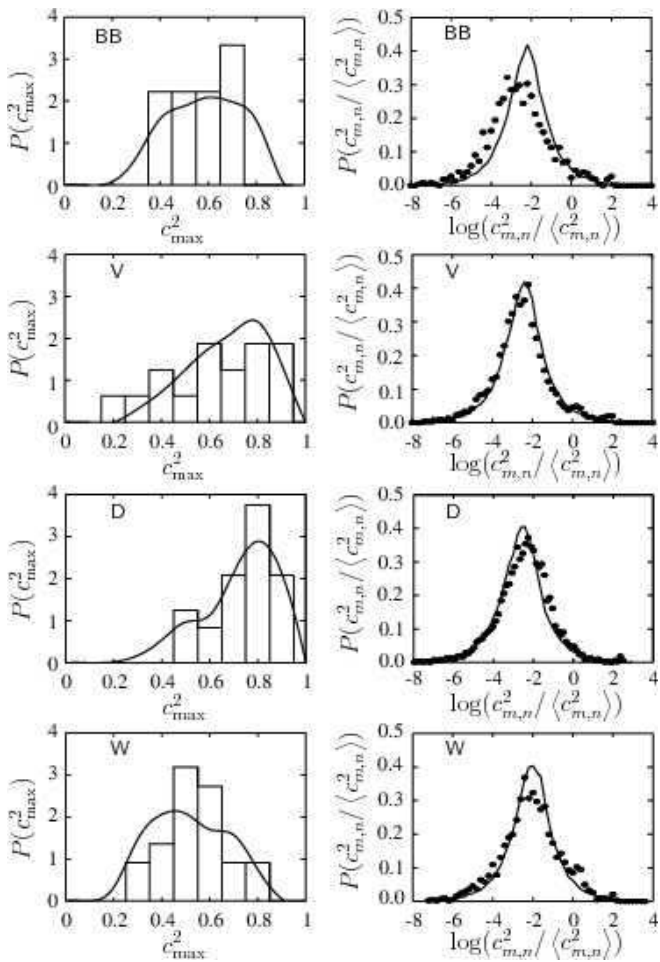}
\caption{Left: $c_{\rm{max}}^2$ distributions of measured superscars
  (histogram), and the fit of the RMT model predictions (solid line).
  Right: normalized distributions of superscar strength spread over
  all states on a logarithmic scale. Experimental distributions (dots)
  are compared with the RMT model predictions.
\label{fig3}}
\end{figure}

As the superscar strength is distributed over a small number of states
only (see Fig.~\ref{fig2}), $v/d$ is smaller than or of the order of
unity. The fit with a BW distribution shows a rather large variation
of the fitted shape as well as the width of the distribution over the
ensemble of observed superscars in a superscar family {\sf F}.  Hence,
the width $\Gamma^{\downarrow}$ is not a well--suited measure to
determine $v/d$. We thus consider the state with the largest coupling
\begin{equation}
c_{\rm{max}}^2={\rm max\,}(c_{m,n}^2)
\end{equation}
for a given superscar which is directly obtained from experiment.
Since a rather small number of states carry strength from the doorway
state (i.e. the constructed superscar), the peak of the fitted
BW shaped strength function usually deviates from the
measured largest superscar strength: The discretely measured state
does not appear exactly at the peak. We may, however, directly compare
the maximal measured value to the corresponding calculated value,
${\rm max\,}(c^2_s (n))$, where $c_{s}$ is obtained from
Eq.~(\ref{cs}).  Not only the average value of $c^2_{\rm{max}}$ can be
studied but also its higher moments. We study the {\em full
  distribution} of these maximal couplings for a superscar family {\sf
  F}, which, as far as we know, has never been considered before. In
Ref.~\cite{Vergini2004} the first two moments of the $c^2_{\rm{max}}$
distribution were studied, but with assumptions not valid in our
context.  The shape of the $c^2_{\rm{max}}$ distribution strongly
depends on the interaction strength, $v/d$, and it is a particularly
sensitive measure for small values of $v/d$, i.e.  of the order one or
smaller.

In Fig.~\ref{fig3} we show measured distributions of $c_{\rm{max}}^2$
with the best fit curves of the RMT model for each superscar family
{\sf F}.  The fit gives the following values for the interaction
strength: for the {\sf BB} superscar $v/d =0.45$, for the {\sf V}
superscar $v/d=0.35$, for the {\sf D} superscar $v/d=0.3$, and for the
{\sf W} superscar $v/d=0.55$.  The coupling strengths are small and
thus our ansatz for a BW shape for the doorway strength function
(Fig.~\ref{fig2}) is in accordance with earlier findings
\cite{Frazier1996}. The {\sf V, D, W} superscar families contain 16,
25 and 22 measured members, respectively, while the {\sf BB} superscar
family contains only 9. The fit in this latter case has thus higher
uncertainty. The averaged measured and calculated $c^2_{\rm{max}}$
values are listed in Tab.~\ref{tab1}.
\begin{table}
\begin{centering}
\vspace{0.5cm}
\begin{tabular}{ccccccc}
\hline \hline
&\multicolumn{3}{c}{\raisebox{0.0ex}[2.3ex]{$\langle c_{\rm max}^2\rangle$}}&\multicolumn{3}{c}{$\Gamma^\downarrow$}\\
\raisebox{1.5ex}[-1.5ex]{{\sf F}}&Exp&RMT&Corr&&Exp&RMT\\
\hline

{\sf BB} &$0.58\pm0.05$&$0.58$&$0.81$&&$0.9\pm0.1$&$1.3$\\

{\sf V} &$0.63\pm0.05$&$0.68$&$0.69$&&$0.8\pm0.1$&$0.8$\\

{\sf D} &$0.74\pm0.03$&$0.72$&$0.69$&&$0.9\pm0.1$&$0.6$\\

{\sf W} &$0.54\pm0.03$&$0.51$&$0.49$&&$1.0\pm0.1$&$1.9$\\
\hline \hline \end{tabular}
\caption{\label{tab1}Experimental results with standard errors of the
  mean versus results from RMT model and directed correlators (Corr)
  for averaged $c_{\rm max}^2$ values and spreading
  width $\Gamma^\downarrow$.}
\end{centering}
\end{table}

Another observable is the distribution of the superscar couplings
over all eigenstates.  The strength of each constructed superscar
is measured (and calculated) over  all 294 states, where the major part
of the strength is concentrated in a few states only.
Figure~\ref{fig3} shows measured distributions compared to
calculations for different interaction strengths obtained from the
fit to the $c^2_{\rm{max}}$ distributions.  Once more, we clearly
see that the model reproduces the experimental distributions for
all superscar families well except in the case of the {\sf BB}
superscar family because of the small number of superscars.

We now turn to the spatial correlations of the wave functions.
Berry~\cite{Berry1977} introduced the correlator
\begin{eqnarray}
C(kr) = \frac{\langle \psi_{k}(\vec{R}-\vec{r}/2)\psi_{k}^*(\vec{R}+\vec{r}/2)
                       \rangle}{\langle |\psi_{k}(\vec{R})|^2\rangle}
\label{corrdef}
\end{eqnarray}
of the wave functions $\psi_{k}(\vec{r})$ where the average is
performed {\it isotropically} over all vectors $\vec{R}$ and, for
fixed moduli of wave vector $\vec{k}$ and $r$, over all directions
of the vector $\vec{r}$. In our context, all wave functions are
real and no complex conjugation is needed in the
definition~(\ref{corrdef}). Berry argued that the spatial
correlations of a wave function in an ergodic system should be
indistinguishable from those of superimposed plane waves.  In two
dimensions this yields the universal prediction $C(kr)=J_0(kr)$,
if possible boundary effects are ignored.  Here $J_0$ is the
Bessel function of order zero. Indeed, this behavior was confirmed
in numerous
systems~\cite{StoeckmannBuch2000,Guhr1998,Doya2002,Schaadt2003}.

\begin{figure}[h]

\includegraphics[width=\linewidth]{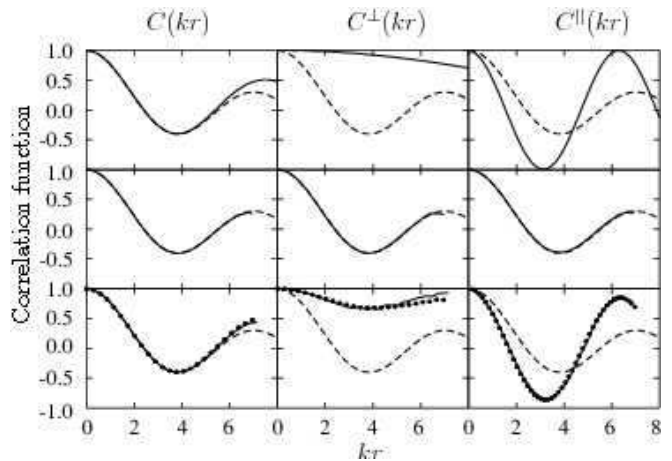}

\caption{The wave function correlators. The $J_0(kr)$ prediction is
  always given as dashed line. The top row shows the correlation
  function of the constructed {\sf V} superscar state as solid lines:
  the isotropic $C(kr)$ as well as the directed $C^{\perp}(kr)$ and
  $C^{||}(kr)$. In the middle row, the same observables are depicted
  as solid lines for the averages over all experimental wave functions
  in the barrier billiard. In the bottom row, the correlators averaged
  over all observed {\sf V} superscars are displayed as solid lines
  and the correlators resulting from Eq.~(\ref{modstate}) with
  $c^2_{\rm{max}}=0.69$ are shown as filled circles.}
\label{fig4}
\end{figure}

The superscars, however, clearly have non--ergodic features.  To
analyze their correlations we define new, especially tailored
observables to which we refer as {\it directed} correlators. Instead
of averaging isotropically as for $C(kr)$, we now carry out the
averages either only across or only along the channel in which the
superscar exists similar to~\cite{Baecker2002}. We thereby obtain the
correlators $C^{\perp}(kr)$ and $C^{||}(kr)$, respectively. In the top
row of Fig.~\ref{fig4} the three correlators of a constructed {\sf V}
superscar $\Psi^{\sf(V)}_{m,n}(\vec{r})$ are depicted.  While the
isotropic correlator $C(kr)$ follows the $J_0(kr)$ prediction up to a
certain scale, the directed correlators strongly deviate from it.  The
results for $C^{\perp}(kr)$ and $C^{||}(kr)$ show that the constructed
superscar fills the channel and moves through it as a sine wave, see
also Fig.~\ref{fig1}.  This information about the form of the waves,
however, is washed out when averaging over all wave functions in the
billiard. As the middle row of Fig.~\ref{fig4} shows, each of the
three correlators worked out for all measured wave functions coincides
with the $J_0(kr)$ prediction for chaotic systems.  Hence, we may use
Berry's random wave approach even though our billiard system is
pseudointegrable.  Importantly we only use the two-point correlations
and only go up to $kr=8$.

We now use these observations to extract information about the
superscar couplings from the measured correlators. Correlators
averaged over all experimentally observed {\sf V} superscars are
displayed in the bottom row of Fig.~\ref{fig4}.  They are similar to,
but slightly different from those for the constructed {\sf V}
superscars in the top row.  The difference is due to the leaking of
the superscar out of the channel or, in the language of the doorway
description, due to the coupling of the background states to the
superscar. We thus model the measured superscars $\Psi^{\sf
  (F)}_{\tilde{f}}(\vec{r})$ for family {\sf F} as a linear
combination of a constructed superscar $\Psi^{\sf
  (F)}_{m,n}(\vec{r})$, which only contributes in the channel, and a
state $\widetilde{\chi}_{k}(\vec{r})$ which is ergodically distributed
everywhere in the billiard,
\begin{equation}
\Psi^{\sf (F)}_{\tilde{f}}(\vec{r}) = c_{\rm max}  \Psi^{\sf (F)}_{m,n}(\vec{r}) +
                   \sqrt{1-c_{\rm max}^2} \widetilde{\chi}_{k}(\vec{r}) \ .
\label{modstate}
\end{equation}
This ansatz is fully consistent with the RMT model set up above and
extends it by also modeling the spatial dependence. The states
describing the background should, first, have $J_0(kr)$ correlations
and, second, be orthogonal to $\Psi^{\sf(F)}_{m,n}(\vec{r})$. Thus, we
choose the ``scarless'' plane waves
\begin{equation}
\widetilde{\chi}_{k}(\vec{r}) =
      \frac{\chi_{k}(\vec{r})-\langle\Psi^{\sf (F)}_{m,n}|\chi_{k}\rangle
                           \Psi^{\sf (F)}_{m,n}(\vec{r})}
             {\sqrt{1-\langle\Psi^{\sf (F)}_{m,n}|\chi_{k}\rangle^2}} \ ,
\label{ergstate}
\end{equation}
with standard plane waves $\chi_{k}(\vec{r})$. The superscar
contribution in the plane waves is small (but not negligible); the
distribution of the overlaps $\langle\Psi^{\sf
  (F)}_{m,n}|\chi_{k}\rangle$ has a standard deviation of $0.13$.  We
convinced ourselves that the correlator of the
$\widetilde{\chi}_{k}(\vec{r})$ follows the $J_0(kr)$ prediction very
closely.  We work out the three correlators for the
model~(\ref{modstate}). They depend on $c_{\rm max}$ which is, just as
in the RMT model above, the coupling to the superscar doorway.  By
fitting to the measured superscar families we determine the couplings
$c_{\rm{max}}$. The fits for the {\sf V} superscar are shown in
Fig.~\ref{fig4}. The resulting $\langle c^2_{\rm{max}}\rangle$ values
in Tab.~\ref{tab1} are close to those obtained from the RMT model.
This is a nice mutual confirmation.  For comparison, we also give the
resulting $\Gamma^\downarrow$ values in Tab.~\ref{tab1}. Obviously,
our new observables are more appropriate.  This is born out in the
large standard deviation of the $\Gamma^\downarrow$ distribution which
is, e.g. for the {\sf W} superscar family, 0.8.

We conclude that our doorway interpretation yields a thorough
understanding of the experimental findings.  Our two new observables
give deeper insight into the statistical features of the doorway
mechanism as such, and it is encouraging to see how well the two
analyses agree.

We thank E. Bogomolny and B. Dietz for valuable discussions.  We
acknowledge support from DFG (SFB 634, SFB/TR12) and from Det
Svenska Vetenskapsr{\aa}det.

\end{document}